\documentclass[prb, twocolumn, amsmath, amssymb, floatfix, showpacs]{revtex4}
\usepackage{color, graphicx}

\begin{document}
\title{Magnetic fields above the surface of a superconductor with 
internal magnetism}
\author{Hendrik Bluhm}
\email{hendrikb@stanford.edu}
\affiliation{Department of Physics, Stanford University, Stanford, CA 94305}

\renewcommand{\vec}[1]{\mathbf{#1}}
\newcommand{\hatv}[1]{\vec{\hat{#1}}}

\begin{abstract}
The author presents a method for calculating the magnetic fields 
near a planar surface of a superconductor with a given intrinsic magnetization
in the London limit. He computes solutions for various magnetic domain 
boundary configurations and derives relations between the spectral densities
of the magnetization and the resulting field in the vacuum half space, which 
are useful if the magnetization can be considered as a statistical quantity and
its features are too small to be resolved individually. The results are useful 
for analyzing and designing magnetic scanning experiments.
Application to existing data from such experiments 
on Sr$_2$RuO$_4$ show that a domain wall would have been 
detectable, but the magnetic field of randomly oriented small domains and 
small defects may have been smaller than the experimental noise level.

\end{abstract}

\pacs{74.25.Ha, 74.20.De, 75.70.-i, 74.70.Pq}
\maketitle

\section{Introduction}

Starting with the discovery of ferromagnetic order in the
superconductors HoMo$_6$S$_8$ \cite{LYNNJW:Dirolr, ISHIKAWAM:Dessbm} 
and ErRh$_4$B$_4$\cite{FERTIGWA:Dessol} about three decades ago,
there has been increasing interest in superconductors with an intrinsic
magnetization. In the above cases and some of the RNi$_2$B$_2$C
compounds (R = rare earth), the magnetization is due to localized
moments of the rare earth ions and coexists with superconductivity in
some temperature range. 
In other materials, the conduction electrons may not only superconduct, but
also carry some magnetization. A prominent example is  
Sr$_2$RuO$_4$ \cite{MAENOY:Suplpw}, which is believed to 
have a complex, time reversal symmetry breaking $p$-wave order parameter,
\cite{MackenzieAP:ThesS2}
so that the orbital angular momentum of the Cooper
pairs creates a magnetic moment.  This is theoretically expected to
cause edge currents at sample boundaries and domain
walls.\cite{MatsumotoM:Quasns, SIGRISTM:Lowmrc} 
A similar effect has been suggested to occur in the $d$-wave superconductor
Na$_x$CoO$_2\cdot y$H$_2$O.\cite{BaskaranG:ElemC2}

The traditional experimental techniques for studying magnetic ordering
phenomena are bulk probes such as muon spin rotation ($\mu$SR) 
or (spin polarized) small angle neutron scattering (SANS).
An alternative approach is to use 
magnetic scanning techniques such scanning Hall probe microscopy (SHPM),
\cite{CHANGAM:ScaHpm} scanning SQUID microscopy,\cite{KirtleyJR:Scasm}
 magnetic force microscopy (MFM) 
\cite{RUGARD:Magfmg,MOSERA:Obssvc} and
magneto-optical techniques.\cite{KoblischkaMR:Magios}
These techniques measure the magnetic field some small distance above 
the surface of the sample as a function of position. 
In many cases, their resolution, 
which is limited by the probe size and probe--sample distance, does not
quite reach the length scales typical for the magnetic structure.
For example, one finds evidence for oscillatory 
magnetic order with a sub-penetration depth length scale in HoMo$_6$S$_8$ 
\cite{LYNNJW:Dirolr, ISHIKAWAM:Dessbm} and ErRh$_4$B$_4$.\cite{FERTIGWA:Dessol}
For larger wave lengths, the Meissner effect in the coexisting state would
screen the field and thus suppress magnetic interactions and destabilize 
the magnetic order. A similar situation has been observed in the
superconductor ErNi$_2$B$_2$C using SHPM.\cite{BluhmH:ScaHpi}
The direct observation of the magnetic fields generated by edge currents
or domain walls in Sr$_2$RuO$_4$ is an ongoing effort which has not produced
any evidence so far\cite{BjornssonPG:ScamiS, KirtleyJ}. 

To plan such scanning experiments and to interpret the resulting data,
it is important to understand to what extent the magnetic field
generated by a spatially varying magnetization inside the sample is
propagated to the probe.  In normal materials, this is a
straightforward magnetostatics problem.  In superconducting samples,
however, it is complicated by the Meissner screening. 
In this work, I present solutions for an infinite planar sample surface
by incorporating the presence of a magnetization $\vec M(\vec x)$ into
a London model and solving the resulting equations, using a 2D
Fourier transform.  This approach follows earlier theoretical
work on superconductors with an internal magnetization due to 
localized magnetic moments,
\cite{GREENSIDEHS:Poscsa, BLOUNTEI:Eleens, NgTK:Spovpd,KUPERCG:Fersvp} 
but should also apply if the magnetism is of different origin, such as a 
spin or orbital moment of the Cooper pairs. To consider the
practically very likely case where the limited measurement resolution
leads to significant averaging over several domains or other features with 
some degree of randomness, I present a
spectral analysis. The resulting relations were employed for the analysis of 
the data in Ref. \onlinecite{BluhmH:ScaHpi} and may also be used to 
analyze null results where no field variation is detectable at the 
experimental noise level. A similar analysis may also be useful for 
interpreting surface sensitive $\mu$SR experiments, which only average over a 
thin layer at the sample surface \cite{MorenzoniE:Nantfi}.

The paper is organized as follows: In section \ref{sec:model}, I
derive the main equations of our model from a generalized
Ginzburg-Landau (GL) functional.  Those equations are solved in a
general framework in section \ref{sec:solution}. In section
\ref{sec:examples}, I discuss simple domain wall and dipole
configurations as examples. Relations between the spectral densities
of the magnetization and the magnetic field in vacuum are computed in
section \ref{sec:spectrum}. Section \ref{sec:SrRu} applies the results to 
recent magnetic scanning work \cite{BjornssonPG:ScamiS} on Sr$_2$RuO$_4$.

\section{Model}
\label{sec:model}
In order to fully describe the interplay between magnetism and 
superconductivity in a phenomenological approach, the 
magnetization $\vec M$ and the superconducting order 
parameter $\psi$ have to be computed self consistently, taking mutual 
interactions into account. This has been done in Refs. 
\onlinecite{GREENSIDEHS:Poscsa, BLOUNTEI:Eleens, NgTK:Spovpd, KUPERCG:Fersvp}
 using the generalized GL functional
\begin{widetext}
\begin{eqnarray}
F[\psi, \vec{M}, \vec{A}] &=& \int d^3 \vec r \left[\frac 1 2 ~ a|\psi|^2 + 
\frac 1 4  b |\psi|^4 + \frac{\hbar^2}{2m^*} ~ \left|\left(i\nabla + \frac{2 e}{\hbar c} ~ \vec{A}\right) \psi\right|^2 \right.
\nonumber \\
&&\left.+\frac 1 2 ~\alpha|\vec{M}|^2 + \frac 1 4 \beta |\vec{M}|^4 
+ \frac 1 2  \gamma^2 |\nabla\vec{M}|^2
+ \frac 1 {8\pi} \vec{B}^2 - \vec{B} \cdot \vec{M}\right], 
\label{eq:GLFunc}
\end{eqnarray}
\end{widetext}
where $\vec B = \nabla \times \vec A$. In this work, I will assume this task 
to be partially solved by starting with a given $\vec M(\vec x)$ and computing
the resulting magnetic field, taking shielding currents into account. 
This approach is clearly justified if the magnetic energy scale is much larger 
than the superconducting one, so that the effect of superconductivity on 
magnetism can be neglected. However, it is also reasonable if the result of 
a self consistent calculation for $\vec M$ is (approximately) known, for 
example from bulk calculations, and one is mainly concerned with the 
effect of the reduced screening at the surface on the observable field.
The errors introduced by this treatment will then primarily  be due to the 
effect of the surface and the modified screening on $\vec M$.

Writing the order parameter as $\psi = |\psi| e^{i\phi}$ and introducing the 
London penetration depth defined by 
$\lambda^{-2} = 4 \pi (|\psi|^2/m^*) e^{*2}/c^2$,
variation of Eq. (\ref{eq:GLFunc}) with 
respect to $\vec{A}$ leads to 
$$\nabla \times (\nabla \times \vec{A} - 4 \pi \vec{M}) + 1/\lambda^2 \vec{A} 
=  (\Phi_0/2\pi \lambda^2)  \nabla \phi.$$
By performing a line integral over a closed loop
after multiplying by $\lambda^2$ and using the Stokes theorem, one obtains
\begin{equation}
\label{eq:varLondon}
\nabla \times (\lambda^2 \nabla \times \vec{B}) + \vec{B} 
= 4 \pi \nabla \times (\lambda^2 \nabla \times \vec{M}) + \Phi_0 \vec f.
\end{equation} 
$\vec f$ is a sum of 2D $\delta$ - functions representing vortex cores, 
which will be ignored in the following.
This result can be obtained directly by treating the supercurrent density
 $\vec{j}_s$ as macroscopic current in the macroscopic Maxwell equation and
thus substituting
$\nabla \times (\vec{B} - 4 \pi \vec{M}) = 4 \pi/c ~ \vec{j}_s$ into 
the London equation 
$\nabla \times (4 \pi \lambda^2/c ~ \vec{j}_s) + \vec{B} = \Phi_0 \vec f$.
Although the above is valid for a spatially varying superfluid density 
$n_s = |\psi|^2/m^*$, I will assume $\lambda$ to be constant,
which leads to the more familiar form 
\begin{equation}
\label{eq:magLondon}
\nabla \times \nabla \times \vec{B} + 1/\lambda^2 \vec{B} 
= 4 \pi \nabla \times  \nabla \times \vec{M}.
\end{equation} 
I assume the magnetic superconductor to occupy the lower half space $z < 0$. 
In  vacuum ($z > 0$), the magnetic field must satisfy 
$\nabla \cdot \vec B = 0$ and $\nabla \times \vec B = 0$. At the interface, 
the normal component of $\vec B$ and the tangential component of 
$\vec H = \vec B - 4 \pi \vec M$ must be continuous. Note that $\vec M$
enters Eq. \ref{eq:magLondon} only through the microscopic current density
$\vec{j}_\vec{M}  = c \nabla \times \vec{M}$. Thus, one may also start
directly from an intrinsic current density $\vec{j}_\vec{M}$ rather than 
$\vec{M}$, which is more natural if an edge current is known from microscopic 
calculations, for example. The appearance of $\vec{M}$
in the tangential boundary condition can also be eliminated by replacing it
with a discontinuity in $\vec{M}$ just below the surface.

\section{Solution}
\label{sec:solution}
In Ref. \onlinecite{KOGANVG:Magfvc}, the field geometry
of a vortex penetrating the surface of an 
anisotropic superconductor for a general orientation of the vortex and the 
main axis of the effective mass tensor with respect to the interface has 
been computed.  In the vortex problem, the right hand 
side appearing in the London equation Eq. (\ref{eq:magLondon}) is a 2D delta 
function instead of the  magnetization term. 
I use the same technique, but only present the calculations for the isotropic 
case for the sake of simplicity.

The Maxwell equations in vacuum can be satisfied by writing the magnetic field 
as $\vec B = -\nabla \Phi$ with $-\nabla^2 \Phi = 0$.
A suitable solution has to be matched to a solution of Eq. (\ref{eq:magLondon})
at z = 0. I solve this problem using a 2D Fourier transform (FT) in 
the $xy$-plane, i.e. by writing a function $A(x, y, z)$ as 
$A(\vec r_{||}, z) = (2 \pi)^{-2} \int d^2\vec k \tilde A(\vec k, z) 
e^{i \vec k \cdot \vec r_{||}}$, with $\vec{r}_{||} = (x, y)$ and 
$\vec{k} = (k_x, k_y)$.
(Note that I use the same symbol for 2 and 3 dimensional vectors, implying 
that the $z$- component vanishes for the latter.)
For $z<0$, the field $\vec B$ is decomposed as $\vec B_0$ + $\vec B_1$.
 $\vec B_0$ is a particular solution  of the inhomogeneous London 
equation (\ref{eq:magLondon}) in full 
 space with boundary conditions at infinity and with the right hand side 
suitably extended to $z>0$. $\vec B_1$ is a general homogeneous solution 
chosen to satisfy the matching condition at the interface.
Under the 2D FT, Eq. (\ref{eq:magLondon}) transforms into 
$(k^2 + 1/\lambda^2 - \partial^2/\partial z^2) \vec{\tilde B}_1 = 0$,
so that
$\vec{\tilde B}_1(\vec k, z) = \vec{B}_{\vec k} e^{K z}$ with 
$K = \sqrt{k^2 + 1/\lambda^2}$.
In vacuum, $-\nabla^2 \Phi = 0$ has the solutions $\tilde \Phi(\vec k, z) = 
\Phi_{\vec k}  e^{-k z}$  so that at $z>0$,
$\vec{\tilde B}(\vec k, z) = (-i \vec k  + k \hatv e_z)  \Phi_{\vec k}  
 e^{-k z}$.

Hence, $\nabla \cdot \vec{B}_1 = 0$ together with  the $\hatv e_z$, $\hatv k$
and $\hatv k \times \hatv e_z$ components of the continuity conditions 
for $\vec B$ and $\vec H = \vec B - 4 \pi \vec M$ at $z = 0$ 
read
\begin{eqnarray*}  
0 & = & i \vec k \cdot \vec B_{\vec k} + K \hatv{e}_z \cdot \vec B_{\vec k} \\
k \Phi_{\vec k} & = & \hatv e_z \cdot 
\left(\vec B_{\vec k} + \vec {\tilde B}_0(\vec k, 0) \right)\\
-i k \Phi_{\vec k} & = & \hatv k  \cdot \left( \vec B_{\vec k} + 
\vec {\tilde B}_0(\vec k, 0) - 4 \pi \vec {\tilde M}(\vec k, 0) \right)\\
0 & = & \left(\hatv k  \times \hatv e_z\right) \cdot 
\left(\vec B_{\vec k} + \vec {\tilde B}_0(\vec k, 0) - 
4 \pi \vec {\tilde M}(\vec k, 0)\right).
\end{eqnarray*}

The last equation for the in-plane transverse component of $\vec B_{\vec k} $
is already decoupled and can be dropped if only the vacuum field is to be 
computed. Solving the first three equations for $\Phi_{\vec k}$ leads to 
\begin{equation}
\label{eq:contCond}
k(K + k) \Phi_{\vec{k}}  =   K \hatv e_z \cdot \vec {\tilde B}_0(\vec{k}, 0)
+ i~ \vec{k} \cdot 
\left(\vec{\tilde B}_0(\vec{k}, 0) - 4 \pi \vec{\tilde M}(\vec{k}, 0)\right)
\end{equation}

If the 2D FT of the inhomogeneous solution, $\vec{\tilde B}_0$, 
cannot be obtained directly, the 3D FT
$\vec B_0(\vec q)$ of $\vec B_0(\vec r)$ can be obtained from $\vec M(\vec q)$
by solving the $3 \times 3$  linear system 
$$-\vec{q}\times (\vec{q} \times \vec{B}_0(\vec q)) +
 1/\lambda^2 \vec{B}_0(\vec q)
=  - 4 \pi\vec{q}\times (\vec{q} \times \vec M(\vec q))$$
 The solenoidal condition $\nabla \cdot \vec B_0 =0$  
will always hold as
$\vec{q} \cdot \vec B(\vec q) = \lambda^2 ~ \vec{q} \cdot 
[\vec{q}\times (\vec{q} \times \vec{B}(\vec q))
- 4 \pi \vec{q}\times (\vec{q} \times \vec{M} (\vec q))] = 0$. 
The 2D FT at $z = 0$ is then  simply 
$\vec{\tilde B}_0(\vec k, 0) = (1/2\pi) \int dq_z \vec B_0(\vec q)$ 
with $\vec q = \vec k +  q_z \hatv e_z$.
While this approach to the inhomogeneous problem is very convenient for 
numerical evaluation and can be generalized to the anisotropic case 
(cf. Ref. \onlinecite{KOGANVG:Magfvc}), 
it is useful to derive an explicit
solution. The component of $\vec M$ parallel to 
$\vec q$ does not contribute to $\vec q \times \vec M$. For  
components of $\vec B_0$ and $\vec M$ orthogonal to $\vec q$,
$\vec q \cdot \vec B_0 = 0$ automatically and the 
vector products simplify to scalar multiplication. 
By decomposing $\vec M$ and $\vec B_0$ into components along the unit vectors
$\hatv e_{xy} = \vec q \times \hatv e_z / |\vec q \times \hatv e_z|$ 
orthogonal to $\vec q$ in the $xy$-plane and 
$\hatv e_\perp = \hatv q \times \hatv e_{xy}$,
the inhomogeneous solution thus simplifies to 
\begin{eqnarray} 
\label{eq:simpleHomSol}
B_{0, \perp} &=& \frac{k^2 + q_z^2} 
{1/\lambda^2 + k^2 + q_z^2}4 \pi M_\perp \nonumber \\
B_{0, xy} &=& \frac{k^2 + q_z^2} 
{1/\lambda^2 + k^2 + q_z^2} 4 \pi M_{xy}
\end{eqnarray}
To evaluate Eq. (\ref{eq:contCond}), the inverse  z-FT must be carried out in 
order to obtain the values at $z = 0$, unless $\vec M$ is $z$- independent.
It turns out that $B_{0,xy}$ does not enter Eq. (\ref{eq:contCond}) because 
of the dot product with $\vec k$. Projecting $B_{0, \perp}$
onto the $z$ and $\vec{\hat k}$ direction, substituting Eq. 
(\ref{eq:simpleHomSol}),
and expressing everything in terms of $M_z$ and $\vec{\hat k} \cdot \vec M$ using 
$M_\perp = \frac{-k}{\sqrt{q_z^2 + k^2}} M_z
+ \frac{q_z}{\sqrt{q_z^2 + k^2}} \vec{\hat k} \cdot \vec M $ leads to
\begin{multline} 
\label{eq:propagation}
k(K + k) \Phi_{\vec k} 
= 2 \int dq_z~ \frac{k^2(K - i  q_z)} 
 {1/\lambda^2 + k^2 + q_z^2} M_z(\vec q)\\
- 2 i\int dq_z~ \frac{K(K - i q_z)} 
 {1/\lambda^2 + k^2 + q_z^2} \vec k \cdot \vec M(\vec q)
\end{multline} 
The $i q_z$ terms in the numerator of the fractions can be dropped if one 
uses the convention that $\vec M$ is extended to $z >0$ as an even function 
so that those terms do not contribute to the integrals.
Eq. (\ref{eq:propagation}) can be summarized qualitatively as follows:
For in-plane components of $\vec M$, the source of $\vec B$ outside the 
superconductor is the divergence of $\vec M$ averaged over one penetration 
depth below the surface. For the normal component, an additional derivative 
is taken, thus increasing the multipole order of the vacuum field by one.
The small $k$ components of the field are just those resulting from  
the subsurface magnetization and its image obtained by reflection about 
a plane $\lambda$ below the surface.

I would like to point out that for the solution method to work as described, 
the interface must be planar and $\lambda^2$ should be constant.
$\vec M$ on the contrary can be an arbitrary function.
However, if $\nabla \times \vec M = 0$ so that $\vec B_0 = 0$ solves Eq.
(\ref{eq:varLondon}), 
the requirement that
$\lambda^2$ may not depend on $z$ can be dropped at the expense of 
solving a more complicated ordinary differential equation instead of the 
Laplace equation to obtain $\vec {\tilde  B}_1(\vec k, z)$. 
For example, discontinuities in $\lambda^2$ as a function of $z$ could be 
treated by matching additional continuity conditions. 
A dependence of $\lambda^2$
on $x$ or $y$ on the other hand would mix different $\vec k$ components and 
thus generally forbid a simple analytic solution.

\section{Examples}
\label{sec:examples}
\subsection{Discussion of Table \ref{tab:examples}}
As examples for various simple, representative
configurations in $\vec M$, I have calculated 
the field of ferromagnetic domain walls, where $\vec M$ changes sign, and 
dipoles at the surface. For $z \gg \lambda$, simple approximate expressions 
in real space can be obtained.  The results are shown in table 
\ref{tab:examples}.
The approximations are based on the fact that
for $z \gg \lambda$,  only $k_x \ll 1/\lambda$  contributes to the Fourier 
integral $B_z(x, z)$. To second order,
$1/(K + |k|) \approx \lambda e^{-|k| \lambda}$
and $K \approx (1/\lambda )(1+ (k\lambda)^2/2)$.
Thus, the approximations  in table \ref{tab:examples}
are good to three and two orders beyond leading order in 
 $k$  for cases (1),(3) and (2), (4), respectively.
In the following, I discuss the far fields obtained 
from these approximations.
To understand those, it is useful to recall that for magnetostatic problems 
in the absence of macroscopic or supercurrents, 
$-\nabla^2 \Phi = - 4 \pi \nabla\cdot \vec M$. Thus, $\nabla\cdot \vec M$
acts as a magnetic charge by analogy with electrostatics.
The field of a discontinuity of the in-plane component of $\vec M$ [case (1)] 
is just twice that of a magnetically charged line with linear
charge density $4 \pi\; 2 M \; \lambda$ situated $\lambda$ below the surface.
It can be understood as 
the charge density due to the discontinuity of $2 M$ in the 
magnetization, which is screened by supercurrents over one penetration depth. 
The additional factor two formally comes from the extension of $M$ to $z > 0$.
For a discontinuity in the normal component $M_z$ [case (2)], one obtains the
dipole field of two lines of opposite magnetic charge with a charge
density $M \lambda$ as above and a separation of $\lambda$. Again, this can be
understood as the screened field of the discontinuity in the
magnetization occurring at the surface.

The localized dipole pointing into or out of the surface [case (3)] has a 
a quadrupole field. An in-plane moment [case (4)] on the contrary has a dipole 
field to leading order.  
For a dipole chain, i.e. $\vec M = \tilde m \delta(x) 
\delta(y) \vec{\hat e}_z$, 
I obtain the same results as for a single dipole oriented in the $z$ direction
[case (3)] with $m = \lambda \tilde m$. The same analogy can be drawn for 
case (4).

It is also of interest to 
consider a configuration where a magnetization $M \hatv e_z$ 
is localized over a width $w$ around $x = 0$. This situation may be 
encountered at an 
antiferromagnetic domain wall, where canting of antiferromagnetically ordered 
in-plane moments produces a local net out-of-plane magnetization. 
The corresponding exact solution is 
a superposition of two solutions for normal discontinuities with opposite 
signs, shifted by $w$. If $w$ is smaller than all other length scales,
$M(x)$ can be replaced by a delta function: 
$M(x) \approx (\int M(x') \; dx') \delta(x) = w M \delta(x)$. The resulting
field is then simply the derivative of that of a discontinuity in $M_z$, i.e.
for $z \gg \lambda$,
\begin{eqnarray*} 
B_z(x, z) &\approx&
- 4 M \lambda^2 w  \frac{d^2}{dx^2} 
\frac{(z + \lambda)}{(z + \lambda)^2 + x^2}
\end{eqnarray*}

\begin{table*}[htb]
\begin{tabular*}{17.9 cm}{|c|c|c|c|}
\hline
Case & $\vec M(\vec r)$  & $B_z$ (exact) & $B_z$ (approximate) \\ 

\hline \hline 
(0) & $M(x)\; \hatv e_y$ & 0 & 0 \\ 
\hline
(1) & $M \mathrm{sgn}(x) \hatv e_x$ &
\parbox{5.7cm}{
$$ - 4 M \int dk~ \frac{1}{|k| + K} e^{i kx} e^{-|k|z}$$} &
\parbox{9.3 cm}{
$$ - 4 M  \int dk~ \lambda e^{ikx} e^{-|k|(z + \lambda)}
 = - 8 M  \lambda  \frac{z + \lambda} {(z+\lambda)^2 + x^2} $$} \\

\hline
(2) & $M  \mathrm{sgn}(x) \hatv e_z$ &
\parbox{5.7cm}{
$$ 4 M \int dk~ \frac{- i k K e^{i kx} e^{-|k|z}}
{(1/\lambda^2 + k^2) (|k| +  K)}$$} &
\parbox{9.3 cm}{ $$4 M \int dk~ 
 (- i k) \lambda^2 e^{ikx} e^{-|k|(z + \lambda)}
= - 8 M \lambda^2 \frac{d}{dx} \frac{z + \lambda}
{(z + \lambda)^2 + x^2}$$}
\\
\hline
(3)& $m \hatv e_z \delta^3(\vec r)$ &
\parbox{5.7 cm}{
$$2 m \int_0^\infty dk~ \frac{k^3}{k + K} J_0(kr) e^{-kz}$$}&
\parbox{9.3 cm}{
 $$ 2 m \int_0^\infty dk~ k^3 \lambda J_0(kx) e^{-k(z + \lambda)}
 =  2 m \lambda \frac{6 (z+\lambda)^3- 9 r^2 (z+\lambda)}
{((z+\lambda)^2 + r^2)^{7/2}}$$}
\\
\hline
(4) & $m \hatv e_x \delta^3(\vec r)$&
\parbox{5.7 cm}{
$$ 2 m \int_0^\infty dk~ \frac{k^2 K} {k+K}  \cos \varphi   J_1(k r) e^{-kz}$$
}& 
\parbox{9.3 cm}{
$$ 2 m \cos \varphi  \int_0^\infty dk~k^2 J_1(k r) e^{-k(z+\lambda)} 
=  2 m \cos \varphi  \frac{3 r (z+\lambda)}{(r^2 + (z+\lambda^2))^{5/2}}$$}\\
\hline

\end{tabular*}
\caption{\label{tab:examples}
Exact and approximate expressions for the magnetic field above the 
superconductor for domain walls and dipoles with an in-plane and out-of-plane 
magnetization. In case (4), $\varphi$ is the angle between the $x$-axis and 
$\vec r$. 
The approximate expressions in the last column are valid for $z \gg \lambda$.
}
\end{table*}

\subsection{Periodic configurations}
If the magnetization is periodic, the Fourier
integrals turn into Fourier sums. If the period $L$ is large 
($L \gtrsim 2 \pi z$), features with size comparable to $L$ can be resolved 
in each unit cell individually and look similar to a solution obtained 
from a constant continuation of $\vec M$ outside that cell.
For shorter periods however, the superposition of many such single cell 
solutions largely cancels out. Formally, this follows from the fact that 
the wave vector $k$
takes only integer multiples of  $2 \pi/L$. Therefore, the dominant 
contributions at $k \lesssim 1/z$ considered in table \ref{tab:examples}
are not present an the leading term becomes that of the lowest 
wave vector $k = 2 \pi/L$. This results in an exponential suppression
by $e^{-2 \pi z/L}$ of the lowest harmonic of the field variation at 
a height $z$ above the surface and all higher harmonics being negligible for
$z \gtrsim L$.

\subsection{Effect of smoothing}
As the exponential cutoff in the inverse FT becomes increasingly sharper 
for larger $z$,
the far field from any feature of finite size in the magnetization
will always be determined by the lowest non-vanishing power of $k$ in 
$ k \Phi_{\vec k}$ for sufficiently large $z$. For a smooth domain wall, 
 $\vec M(\vec q)$ has less weight at large $\vec q$  compared to a 
sharp discontinuity of equal magnitude, but the values at small $\vec q$ 
are affected little.
Thus, the asymptotic results for sharp discontinuities in $\vec M$ 
are still valid if $\vec M$ changes smoothly over a width $w$ as long as 
$z \gg w$. For example, the far field of a single domain
boundary does not depend on the length scale over which the
magnetization changes in the $xy$-plane, but only on the difference
between the asymptotic values of $M$ on both sides. Hence, the approximate 
results in table \ref{tab:examples}  are of rather general
validity. For a graphic illustration, see Fig. \ref{fig:domainfield}, which 
is discussed in  Sec. \ref{sec:SrRu}.
 The situation is slightly different for a nontrivial $z$-dependence of 
$\vec M$: the $q_z$-cutoff is always determined by the 
$1/(1 +(\lambda q_z)^2)$ 
factor for small $k$. In real space, this corresponds to an exponentially 
weighted average over a distance of $\lambda$ below the surface.

\section{Spectral Analysis}
\label{sec:spectrum}

The resolution of currently available  magnetic imaging techniques
is often not sufficient to resolve an actual domain structure. 
In practice, averaging occurs both due to the imaging 
height $z$ and the finite sensor size. Here, I will only consider the more 
universal height effect.
For a height larger than the typical domain size, the magnetic field represents
an average over several domains. As shown in the previous section, this 
would lead to an exponential suppression for periodic configurations.
However, domains will usually not be strictly periodic, but have some 
distribution of size. Therefore, a statistical description
is most adequate. Assuming a given correlation function and thus spectrum 
of the magnetization $\vec M(\vec r)$, I compute the spectrum of the 
resulting magnetic field. 

The spectral function 
of two functions $f(\vec r)$ and $g(\vec r)$ in $d$ dimensions,
$S_{fg}(\vec q) =  \int d^d\vec{r'} 
\langle f(\vec r)~g(\vec r + \vec r') \rangle 
e^{-i \vec r' \cdot \vec q}$, satisfies the relation  
$\langle f(\vec q)^*  g(\vec q') \rangle 
= (2 \pi)^d \delta^d(\vec q - \vec q') S_{fg}(\vec q)$
Here, $\langle \cdot \rangle$ stands for an ensemble average over different 
realizations of $f$ and $g$. It is implicit to this definition that the 
correlator $\langle f(\vec r)~g(\vec r + \vec r') \rangle$ is independent of
$\vec r$. 

It is convenient to introduce the propagation coefficients
\begin{eqnarray}
  c_{\vec q, z} &=& 2k^2 \lambda^2 \gamma_{\vec q}
\nonumber  \\
c_{\vec q, \alpha} &=& 
-2ik _\alpha K \lambda^2 \gamma_{\vec q} \quad (\alpha = x,y) \\
\gamma_{\vec q} &=&  \frac{(K - i q_z)e^{-k z}} 
 {\lambda^2(k + K)(1/\lambda^2 + k^2 + q_z^2)}
\end{eqnarray}
and rewrite  Eq. (\ref{eq:propagation}), using 
$\tilde B_z(\vec k, z) = k \Phi_{\vec k} e^{-k z}$, as
$B_z(z_0, \vec k) = \int dq_z \sum_\alpha c_{\vec q, \alpha} M_\alpha(\vec q)$.
It follows that 
\begin{eqnarray}
S_{B_z}(\vec k, z) &=& 2\pi \int dq_z \sum_{\alpha,\beta} 
c_{\vec q, \alpha}^* c_{\vec q, \beta} S_{M,\alpha\beta}(\vec q)\\
\label{eq:spect}
&=& 2\pi \int dq_z \sum_\alpha 
|c_{\vec q, \alpha}|^2 S_{M,\alpha\alpha}(\vec q) \nonumber\\
&+& 4\pi \int dq_z \sum_{\alpha \neq \beta} 
\mathrm{Re}(c_{\vec q, \alpha}^* c_{\vec q, \beta} S_{M,\alpha\beta}(\vec q)).
\end{eqnarray}
I have used $S_{M, \alpha \beta}$ as a short hand notation for the
spectral function  $S_{M \alpha M \beta}$ of two different components of 
$\vec M$.
Similar expressions can be written down for spectral densities involving 
other components of $\vec B$.

Because of the presence of the surface, the assumption of
translational invariance in the $z$-direction  implicit to the definition 
of $S_M$ is by no means
trivial. If the presence of the surface does not affect the
structure of $\vec M$ too much, and the range of the surface influence is
much shorter than $\lambda$, most of it should average out
because $B_z$ is sensitive to what happens within a layer of thickness
$\lambda$ below the surface.
However, the $z$-invariance is only required in order to define 
spectral functions $S_{M, \alpha \beta}$. 
If the interface (or other effects) do break the $z$-invariance of 
$\langle  M_\alpha(\vec r)~ M_\beta(\vec r + \vec r') \rangle$
so that it depends on both $z$  and $z'$,
it is still possible to derive an expression for $S_{B_z}$
similar to  Eq. (\ref{eq:spect}), however involving a double 
integral over $q_z$.

\begin{figure}
\includegraphics{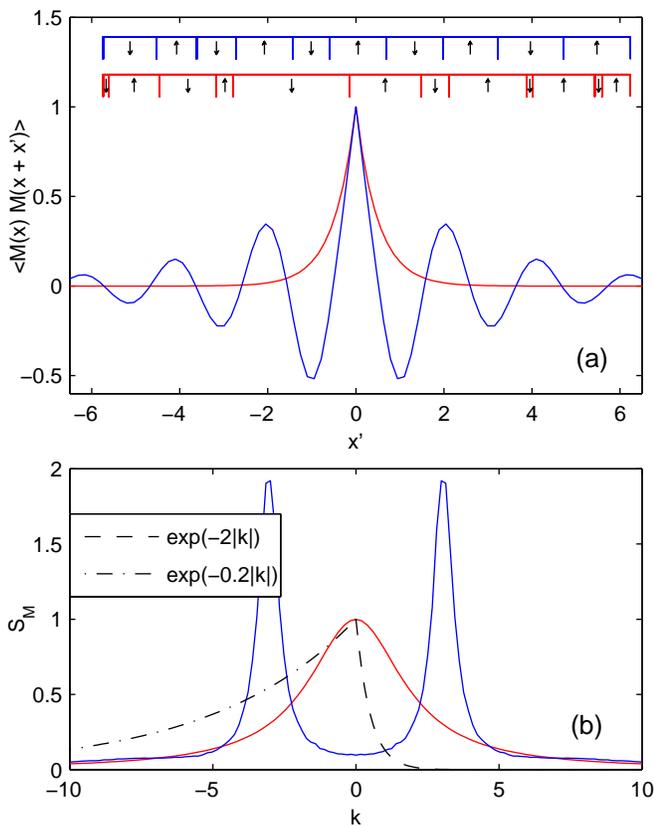}
\caption{\label{fig:corel} (Color online)
Illustration of different one-dimensional domain structures for $|M| = 1$.
(a) Correlator of $M$, and small sample of the corresponding  
 real space domain structure (inset). (b) Corresponding spectral functions. 
The domain width in the red (lower) structure is exponentially distributed
with mean 1, which gives  $\langle M(x) M(x + x')\rangle = e^{-2 |x'|}$ and
$S_M(k) = 1/(1 + k^2/4)$.
The domain width in the blue (upper) structure has a  
Gaussian distribution with mean 1 and standard deviation 0.3, leading to 
an oscillatory correlator and peaks in the spectral function at $k =\pm \pi$. 
The curves for this case were 
obtained numerically from an ensemble of 10$^5$ domains. 
The dashed lines in (b) represent the propagation factors $e^{-|k|z}$ 
for heights $z = 2$ and $z = 0.2$. For large $z$, 
only $S_M(k\approx 0)$ is relevant.
}
\end{figure}

A statistical analysis will be  most relevant when 
the measurement height $z$ is much larger than any of the intrinsic 
length scales of the variation of $\vec M$, 
i.e. too large to resolve individual features.
In this case, $S_M(\vec k + q_z \vec {\hat e}_z)$ will not have a strong 
$\vec k$  dependence in the small $\vec k$ region 
surviving the $e^{-kz}$ cutoff 
and can be approximated by $S_M(q_z \vec {\hat e}_z)$. 
For $z \gg \lambda$, a similar approximation can be made for the propagation
coefficients $c_{\vec q, \alpha}$ and $\gamma_{\vec q}$:
\begin{eqnarray}
 c_{\vec q,\alpha} &\approx& -2ik_\alpha \lambda \gamma_{\vec q}\quad (\alpha = x, y)\\
 |\gamma_{\vec q}|^2 &\approx&  \frac{e^{-2k (z + \lambda)}}
{1+ q_z^2 \lambda^2} 
\end{eqnarray}
In any case, the properties of $\vec M$ only enter via the integrals 
\begin{equation}
\label{eq:intqz}
\int dq_z \frac{1}{1 + k^2\lambda^2 + q_z^2\lambda^2} S_{M, \alpha \beta}(\vec q)
\end{equation}
As argued above, it will often be a good approximation to set $\vec k=0$.
In many cases, $S_M(q_z \vec{\hat e}_z)$ will have a peak at some wave vector
$q_0$ (and consequently at $-q_0$), similar to the illustration in Fig. 
\ref{fig:corel}. 
For simplicity, I assume that there is only one such maximum.
A finite $q_0$ is a signature of an oscillatory behavior of $\vec M$. 
The width of the peak corresponds to the inverse coherence length of the 
oscillation or the correlation length for $q_0 = 0$.

The integral (\ref{eq:intqz}) can be approximated further in two limiting 
cases. If the 
coherence length of $\vec M$ along the $z$ direction is much larger 
than $\lambda$, then $S_M(q_z \vec{\hat e}_z)$ is sharply peaked, and
the kernel $1/(1 + q_z^2\lambda^2)$ can be replaced by  $
1/(1 + q_0^2\lambda^2)$ and pulled out of the integral.
In the opposite limit, the peak in $S_M(q_z \vec{\hat e}_z)$ is much wider than
$1/\lambda$ so that the $q_z$ dependence of $S_M$ can be  
neglected entirely and one obtains 
$$\int dq_z \frac{1}{1 + q_z^2 \lambda^2} S_{M, \alpha \beta}(\vec q)
\approx (\pi/\lambda) S_M(0)$$
Assuming that the $\alpha$-component  of $\vec M$ dominates, the respective
diagonal term of Eq. (\ref{eq:spect}) takes the form
$$S_{B_z}(\vec k) = (8\pi^2/\lambda) (k_\alpha \lambda)^2 
e^{-2k(z + \lambda)} S_{M, \alpha \alpha}(0)$$
for $\alpha = x, y$.
If the $M_z$ component is dominant, $(k_\alpha \lambda)^2$ in the prefactor 
must be replaced by $(k\lambda)^4$.
Similar expressions can be written down for the off-diagonal components
of $S_{M, \alpha\beta}$. 
For an illustration of the relation between a measured $B_z$ and its spectral
function, the reader is referred to Ref. \onlinecite{BluhmH:ScaHpi}.
Integration leads to simple expressions for 
$\langle B_z^2 \rangle = (2\pi)^{-2} \int d^2 \vec k S_{B_z}(\vec k)$:
 \begin{eqnarray}
\langle B_z^2 \rangle &=& \frac{3 \pi} 4 \frac \lambda {(z+\lambda)^4} 
S_{M, \alpha\alpha}(0)  \quad \text{for } \alpha = x, y\\
\label{eq:randz}
\langle B_z^2 \rangle &=& \frac {15\pi} 2 \frac {\lambda^3}{(z+\lambda)^6}
 S_{M, zz}(0).
\end{eqnarray}
Those expressions can be used to estimate the signal expected in a scanning
experiment or to estimate $S_M(0)$ from the observed field variation. 
Note that if the variation of the domain size is sufficiently large for
$\langle \vec M(\vec r) \vec M(\vec r + \vec r')\rangle$ to be essentially 
non-negative, as for the exponential width distribution in Fig. 
\ref{fig:corel}, $S_{M, \alpha \alpha}(0) 
= \int d^3 \vec r' \langle M_\alpha(\vec r) M_\alpha(\vec r + \vec r')\rangle$
can be interpreted as the product of a correlation volume 
$\int d^3 \vec r' \langle M_\alpha(\vec r) M_\alpha(\vec r + \vec r')\rangle 
/ \langle M_\alpha^2 \rangle$ and the mean square magnetization  
$\langle M_\alpha^2 \rangle$.

\section{Application to S\lowercase{r}$_2$R\lowercase{u}O$_4$}
\label{sec:SrRu}
Based on various evidence, it is believed that Sr$_2$RuO$_4$ is a
spin triplet superconductor with a $p$-wave order parameter of the same 
symmetry class as $k_x \pm i k_y$.\cite{MackenzieAP:ThesS2, NelsonKD:OddsS2}
Convincing evidence that the order parameter is indeed time-reversal symmetry 
breaking (TRSB) has recently been obtained by Sagnac-interferometry 
experiments.\cite{XiaJing:HigrpK}
Such a TRSB order parameter is expected to cause  chiral  
currents at sample edges, domain walls or impurities.
\cite{MatsumotoM:Quasns,SIGRISTM:Lowmrc}
The direct observation of such effects in  Sr$_2$RuO$_4$ is an 
ongoing effort. 
So far, the most direct indication of spontaneous fields is given by $\mu$SR 
data, reporting  ``a broad distribution of fields 
arising from a dilute distribution of sources".\cite{LukeGM:TimssS}
Phase sensitive tunneling measurements 
support the notion of small chiral domains.\cite{KidwingiraF:Dynsop} 
Scanning Hall probe and scanning SQUID microscopy experiments
\cite{BjornssonPG:ScamiS, KirtleyJ, DolocanVO:Obsvca} on the other hand did 
not detect any sign of a spontaneous magnetization associated with 
superconductivity. 

The magnetic scans of the $ab$-face 
in Ref. \onlinecite{BjornssonPG:ScamiS} showed neither 
localized features nor a random field variation
that could be attributed to TRSB. Even holes that were drilled using a focused
ion beam (FIB) failed to show a magnetic signature. Lacking suitable 
theoretical models, a quantitative analysis of those null results was 
inconclusive.
In this section, I will use the results derived above to compute the expected
field from domains and defects. This allows to set certain limits on
the internal magnetization strength that would be consistent with the data.

A complete description of chiral domains requires a self consistent
computation of the order parameter and magnetic fields. This has been
carried out using microscopic theory \cite{MatsumotoM:Quasns} 
and a GL approach\cite{SIGRISTM:Lowmrc} without considering the effect of
a surface. Such detailed calculations generally require a numerical
solution. Taking the presence of a surface into account leads to a
further complication. Thus, they are rather cumbersome for the purpose
of data analysis and planning experiments. 

For a domain wall along the $x = 0$ plane in an infinite sample, 
the results of those computations generally show a current along
the $y$-direction which decays over about one in-plane coherence length 
$\xi_{ab}$ in the $x$-direction and changes sign. The counterflowing current
decays on the scale of $\lambda$ such that the magnetic field far
inside each domain vanishes. Indeed, one can obtain good fits of the form
\begin{equation}\label{eq:matsufit}
B_{0,z}(x) = \frac{B_0}{1 - \tilde \xi^2/\tilde \lambda^2} 
\mathrm{sgn}(x)(e^{-|x|/\tilde \lambda} 
- e^{-|x|/\tilde \xi})
\end{equation}
to the numerical results for the magnetic 
field of Ref. \onlinecite{MatsumotoM:Quasns}, 
with  $\tilde \lambda = 2.2~\xi_{ab}$, $\tilde \xi = 1.5~\xi_{ab}$, 
and $B_0 = 87$ G. I used  $\xi_{ab}$ = 66 nm and $\lambda$ = 150 nm
\cite{MackenzieAP:ThesS2} to compute the thermodynamic critical field
entering the prefactor of the result of Ref. \onlinecite{MatsumotoM:Quasns}. 
It is easy to show that this expression is a solution
to the London equations for $\lambda = \tilde \lambda$ in the presence of a
chiral current density $j_y(x) = -(cB_0/4 \pi \tilde \xi) e^{-|x|/\tilde \xi}$.
This current can be identified with an internal magnetization 
$M_z = (B_0/4\pi) \mathrm{sgn}(x) (1- e^{-|x|/\tilde \xi})$ in the 
$z$ direction due to the orbital magnetic moments of the Cooper pairs. 
The value of $\tilde \lambda$ is in good agreement with the 
value of $\kappa \equiv \lambda/\xi_{ab} = 2.5$ assumed in Ref. 
\onlinecite{MatsumotoM:Quasns}.
This suggests that the London approach captures screening effects quite
accurately, and therefore should give a good approximation for the field 
above a surface. Of course, this will neglect some features in the 
full GL solutions. For example, the latter 
show a slight depression of the superfluid
density near the domain wall, which is neglected here by assuming a constant 
$\lambda$. Similar modifications of both the superfluid density and 
the chiral current density at the surface should be small due to the short 
$c$-axis coherence length $\xi_c = 0.05~\xi_{ab}$. The anisotropy is of no 
consequence because all currents flow along the $ab$-plane so that only 
$\lambda_{ab}$ matters. 
 
\begin{figure}[htb]
\includegraphics{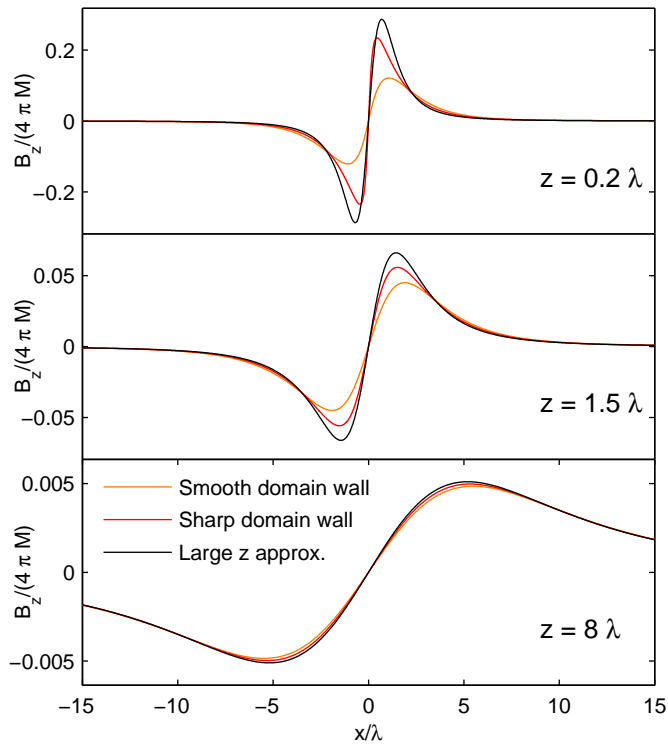}
\caption{\label{fig:domainfield}(Color online)
Field profile at different heights $z$ above a domain wall were 
$M_z$ changes sign.
The curves labeled ``smooth'' and ``sharp domain wall'' were obtained from the
full solution with $M_z(x) = M \mathrm{sgn}(x)(1 - e^{-|x|/(0.7 \lambda)})$
and $M_z(x) = M \mathrm{sgn}(x)$ respectively. The latter solution and the 
corresponding large $z$ approximation as shown are given by table 
\ref{tab:examples}, case (2). The value of the domain boundary 
width was chosen according to the fit to the numerical results 
for Sr$_2$RuO$_4$ from Ref. \onlinecite{MatsumotoM:Quasns}.
$z = 8 \lambda$ corresponds to the Hall probe scans from Ref. 
\onlinecite{BjornssonPG:ScamiS}.
}
\end{figure}

Since the far field of the domain wall only depends on the difference
of the asymptotic values of $M$ away from the domain wall, 
and experimentally $\xi_{ab} = 66\text{ nm} < \lambda = 150 
\text{ nm} \ll z \approx 1~\mu$m, it is appropriate to use the large-$z$ 
results for a sharp discontinuity in $M$ to analyze the results of Ref. 
\onlinecite{BjornssonPG:ScamiS}.
If accurate results at a $z \lessapprox 3 \lambda$ are required, 
Eq. (\ref{eq:contCond}) together with the above approximation for 
$B_{0,z}(x)$ should be used. Fig. \ref{fig:domainfield} shows the 
field profiles for those two methods and the exact result for a sharp domain 
wall at different heights.

In the experiments, the rms noise levels were 35 mG and 0.45 mG at 
imaging heights $z = 1.2~\mu$m and 2 $\mu$m in the Hall probe and
SQUID scans, respectively.
 Correction factors due to oversampling and averaging 
over the size of the  SQUID pickup loop are of order unity. 
An isolated domain wall would result in a field of 
$5 \cdot 10^{-3} 4 \pi M = 0.4$ G and
 $2 \cdot10^{-3} 4 \pi M = 0.17$ G respectively at the above imaging 
heights and $B_0 = 87$ G, and should be clearly visible in the data.
Thus, there was either no such domain wall in the scanned area, or its
magnetization was $4 \pi M < 7$ G and $4 \pi M < 0.2$ G respectively,
so that it was hidden by sensor noise.

The exact calculation of the signature of a hole is more difficult
because the translational invariance of the boundary conditions is
broken. However, if the diameter of the hole or defect, which I assume
to extend along the $z$-direction normal to the surface, is much
smaller than $\lambda$, the absence of superfluid in it can be
neglected and the dipole calculation should be a good approximation.
For a hole or defect with a volume of $\xi_{ab}^3$, the maximum field
according to case (3) in table \ref{tab:examples} at $z = 1.2~ \mu$m
and 2 $\mu$m is $1.2 \cdot 10^{-5} 4 \pi M = 1.1$ mG and $2 \cdot 10^{-6}
4 \pi M = 0.17$ mG, respectively. 
 This signal would be nearly undetectable at the experimental
noise level of Ref. \onlinecite{BjornssonPG:ScamiS}. Furthermore,
the extent of the defect along the $z$-direction could be as small as
$\xi_c = 0.05~\xi_{ab}$, and the order parameter is not necessarily
suppressed entirely.  For a columnar defect on the other hand, one
factor of $\xi_{ab}$ has to be replaced by $\lambda$, decreasing the
limit on $M$ only by about a factor three.  Since the FIB drilled
holes in the experiment were significantly larger (about 1 $\mu$m),
they have both a larger moment and less effective Meissner screening.
This leads to a stronger signal whose calculation goes beyond
the scope of this paper.

One can also estimate the signal expected from a random configuration of small
domains. The smallest conceivable domain volume is on the 
order of $\xi_{ab}^2 \xi_c$.
Assuming that the domain size fluctuates enough to use this as 
correlation volume, Eq. (\ref{eq:randz})
implies that the rms signal could be as small as 0.7 mG and 0.16 mG, again
less than  the experimental noise.
A domain size distribution that does not  satisfy the assumptions leading
to  Eq. (\ref{eq:randz}) may result in even smaller signals.
Thus, the possibility of very small domains cannot be ruled out. 

In all the above cases, the smaller imaging height of the Hall probe 
does not compensate for its large noise compared to the SQUID. 
Assuming the predicted magnitude of the chiral currents 
\cite{MatsumotoM:Quasns} is correct, the calculations
show that any domain wall should have been detected by the measurements.
Small defects on the other hand might easily have been hidden in the noise.
It is also possible that a random signal from domains would have been 
too small to observe, especially if the domains are short in the $c$-direction
or very homogeneous in size while not too large. However, it appears that one 
should not take the notion of a magnetization due to $p$-wave pairing too 
literally. It was shown \cite{BraudeV.:Orbmdc} that the chiral currents can in 
general {\em not} be written as the curl of a global magnetization, 
and that the effective value of $\vec M$ depends on the type of domain wall.
				   
Note that a similar analysis of the magnetic scanning data of Refs. 
\onlinecite{BjornssonPG:ScamiS, KirtleyJ}, also making use of the relations 
derived in Sec. \ref{sec:solution}, has been carried out 
in parallel with the present work in Ref. \onlinecite{KirtleyJ}.

\section{Conclusion}
I have presented a model for a superconductor with an intrinsic 
magnetization by combining the macroscopic magnetostatic 
Maxwell equation with the London 
relation and obtained the field geometry at a planar
superconductor-vacuum interface for a given spatial variation of the 
magnetization.
Solutions for a range of specific magnetic domain boundary configurations
give simple expressions in the limit of a large height above the sample.
If the height above the surface at which the magnetic field can be measured 
exceeds the characteristic length scale of variations in the magnetization,
a spectral analysis can be used to relate the spectral densities of the two 
at resolvable wave vectors.
If a specific model for the structure of the magnetization is at hand, a direct
comparison with the measured field is possible. Otherwise, some simplifying 
assumptions give a simple estimate relating the spectral density 
of $\vec M$ at the superconductor - vacuum interface to the observable 
spatial rms-variation 
of the magnetic field. As an example for an application, I have applied my 
calculations to recent experimental 
results on Sr$_2$RuO$_4$\cite{BjornssonPG:ScamiS}, concluding that large 
chiral domains would have been visible in those experiments, but small domains
and defects may have escaped detection.

\acknowledgments{I would like to thank Kam Moler, Ophir Auslaender and 
John Kirtley for giving feedback on the manuscript.  
This work has been supported by the Department of
Energy under contract DE-AC02-76SF00515.}

\bibliography{bibdata_ENBC,bibdata_magsc}

\end{document}